\documentclass[aps,prl,superscriptaddress,reprint,longbibliography]{revtex4-1}
\usepackage{graphicx}
\usepackage{amsmath}
\usepackage{times}
\usepackage{amsfonts}
\usepackage{amssymb}
\usepackage[colorlinks=true,linkcolor=blue,citecolor=blue,urlcolor=blue]{hyperref}

\renewcommand{\Im}{\operatorname{Im}}

\renewcommand{\eqref}[1]{Eq.~(\ref{eq:#1})}
\newcommand{\eqreftwo}[2]{Eqs.~(\ref{eq:#1},\ref{eq:#2})}
\newcommand{\eqrefrange}[2]{Eqs.~(\ref{eq:#1}--\ref{eq:#2})}

\newcommand{\figref}[1]{Fig.~\ref{fig:#1}}
\newcommand{\cc}[1]{\overline{#1}}

\newcommand{\vect}[1]{\mathbf{#1}}
\newcommand{\tens}[1]{\boldsymbol{#1}}

\begin{document}

\title{Shape-independent limits to near-field spectral radiative heat transfer}

\author{Owen D. Miller}
\affiliation{Department of Mathematics, Massachusetts Institute of Technology, Cambridge, MA 02139}
\author{Steven G. Johnson}
\affiliation{Department of Mathematics, Massachusetts Institute of Technology, Cambridge, MA 02139}
\author{Alejandro W. Rodriguez}
\affiliation{Department of Electrical Engineering, Princeton University, Princeton, NJ 08544}

\begin{abstract}
    We derive shape-independent limits to the spectral radiative heat-transfer rate between two closely spaced bodies, generalizing the concept of a black body to the case of near-field energy transfer. Through conservation of energy and reciprocity, we show that each body of susceptibility $\chi$ can emit and absorb radiation at enhanced rates bounded by $|\chi|^2 / \Im \chi$, optimally mediated by near-field photon transfer proportional to $1/d^2$ across a separation distance $d$. Dipole--dipole and dipole--plate structures approach restricted versions of the limit, but common large-area structures do not exhibit the material enhancement factor and thus fall short of the general limit. By contrast, we find that particle arrays interacting in an idealized Born approximation (i.e., neglecting multiple scattering) exhibit both enhancement factors, suggesting the possibility of orders-of-magnitude improvement beyond previous designs and the potential for radiative heat transfer to be comparable to conductive heat transfer through air at room temperature, and significantly greater at higher temperatures.
\end{abstract}

\maketitle

\date{\today}

Heat exchange mediated by photons, or radiative heat transfer, can be dramatically modified for bodies separated by small gaps~\cite{Polder1971,Rytov1988,Mulet2002,Joulain2005,Volokitin2007,Rousseau2009,Basu2009}. We exploit energy-conservation and reciprocity principles to derive fundamental limits to the near-field spectral heat flux between closely spaced bodies of arbitrary shape, given only their material susceptibilities $\chi(\omega)$ and their separation distance $d$. Our approach enables us to define optimal absorbers and emitters in the near field, which contrast sharply with far-field black bodies: their response is bounded by the amplitude of their volume polarization currents, rather than their surface absorptivities, and maximum energy transfer requires coordinated design of the two bodies (whereas the far-field limit derives from the properties of a single black body). These distinguishing characteristics lead to two possible enhancements relative to black-body emission: a material enhancement factor $|\chi(\omega)|^2 / \Im \chi(\omega)$ that represents the maximum absorber and emitter polarization currents, and a near-field enhancement factor $1 / d^2$ that represents maximum interaction between currents in free space. We show that restricted versions of our limits can be approached for sphere--sphere and sphere--plate configurations. For two extended structures, however, common planar geometries---including bulk metals~\cite{Loomis1994,Xu1994,Pendry1999,Fu2006,Hu2008,Ottens2011,VanZwol2011,Basu2009a,Ben-Abdallah2010a,Nefzaoui2013}, metamaterials~\cite{Biehs2011,Francoeur2011,Joulain2010,Biehs2012,Biehs2013,Guo2013,Narimanov2012}, and thin films~\cite{Biehs2007,Francoeur2008,Francoeur2009,Ben-Abdallah2009,Francoeur2010,Basu2011,Miller2014a}---exhibit flux rates orders of magnitude short of the limits because they do not satisfy the optimal-absorber condition. Instead, we find that idealized plasmonic-particle arrays, interacting within a Born approximation with negligible multiple scattering, approach the limits at selected frequencies, and that the possibility of reaching the limits, even over a narrow bandwidth (a desirable feature for thermophotovoltaics~\cite{Whale2002,Laroche2006,Basu2009,Bermel2010,Rodriguez2011}), would represent an orders-of-magnitude improvement over current designs.

A ray-optical black body absorbs every photon incident upon its surface, which by reciprocity (Kirchoff's Law) yields its emissivity and the black-body limit to thermal radiation~\cite{LienhardIV2011}. At wavelength and subwavelength scales, nanostructures can exhibit optical cross-sections much larger than their physical cross-sections~\cite{Bohren1983}, making it difficult even to define quantities like emissivity. A further difficulty in the near field is the presence of evanescent waves, which can increase transmitted power but only through interference with reflected waves~\cite{Johnson2002a}. Although the possibility of enhancement beyond the blackbody limit was realized by Rytov, Polder, and others in the 1950s~\cite{Polder1971,Rytov1988}, efforts to find underlying limits have been restricted to planar structures with translation symmetry (including metamaterials), without consideration of material loss~\cite{Pendry1999,Basu2009a,Ben-Abdallah2010a,Basu2011,Biehs2012,Nefzaoui2013}. Spherical-harmonic~\cite{Pozar2009,Liberal2014} and Green's-function~\cite{Hugonin2015} limits are difficult to apply in the near field where a large but unknown number of spherical harmonics can be excited by general shapes~\cite{Miller2015subm}.

Without reference to particular structures or symmetries, assuming only linear electromagnetism, we translate the reciprocity principle to the near field by applying it to polarization currents within the bodies. Dipoles in vacuum exchange energy at a rate limited by the energy density of an outgoing free-space wave~\cite{Miller2000a}. As we show below, the maximum energy transfer between material bodies occurs when the currents within the bodies couple individually at the dipole--dipole limit, amplified by material enhancement factors. These conditions allow for much greater heat transfer than has previously been shown possible. 

\begin{figure*}[t!]
    \includegraphics[width=\linewidth]{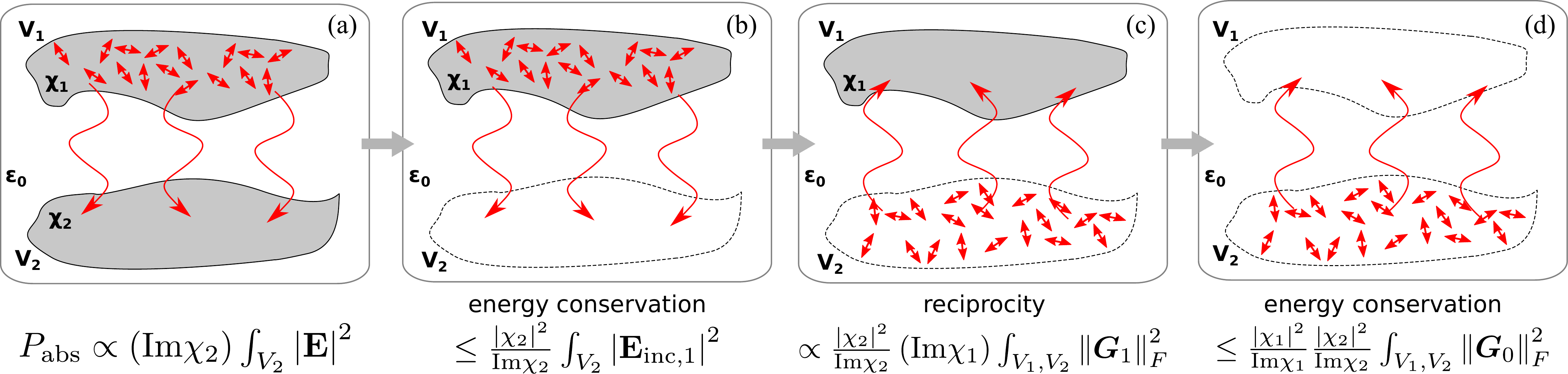}
    \caption{\label{fig:fig1}(a) Radiative heat transfer: Fluctuating currents in an emitter (body 1, susceptibility $\chi_1)$ generate a field $\vect{E}_{\textrm{inc},1}$ and transfer energy to an absorber (body 2, susceptibility $\chi_2$) at a rate $P_{\textrm{abs},2}$. (b) Energy conservation bounds $P_{\textrm{abs},2}$ in terms of $\vect{E}_{\textrm{inc},1}$, and a resonant enhancement factor $|\chi_2|^2 / \Im \chi_2$. (c) The sources and ``receivers'' can be exchanged by reciprocity, whereupon (d) absorption in body 1 is bounded, yielding a spectral-flux limit determined by $\chi_1$, $\chi_2$, and the free-space GF $\tens{G}_0$. For near-field transfer the GF integral is $\sim1/d^2$, for separation $d$.}
\end{figure*}
Radiative heat exchange is depicted schematically in \figref{fig1}(a): fluctuating currents arise in body 1 at temperature $T_1$, and transfer energy to body 2 at a rate of~\cite{Joulain2005}
\begin{align}
    H_{1 \rightarrow 2} = \int_0^{\infty} \Phi(\omega) \left[\Theta(\omega,T_1) - \Theta(\omega,T_2)\right] \mathrm{d}\omega,
    \label{eq:heat_transfer}
\end{align}
where $\Phi(\omega)$ is a temperature-independent energy flux and $\Theta$ is the Planck spectrum. $\Phi(\omega)$ is the designable quantity of interest, to be tailored as a function of frequency depending on the application and available materials.

\emph{Limits---}The spectral heat flux $\Phi(\omega)$ is the power absorbed in body 2 from fluctuating sources in body 1 (or vice versa). In recent work~\cite{Miller2015subm} we have bounded the scattering properties of any dissipative medium excited by a known, externally generated incident field. The bounds arise from the functional dependencies of power expressions with respect to induced currents: absorption is a quadratic functional, whereas extinction (absorption+scattering), given by the optical theorem~\cite{Jackson1999,Newton1976,Lytle2005,Hashemi2012}, is only a linear functional. Energy conservation requires that extinction be greater than absorption, which imposes a bound on the magnitude of the excited currents. Radiative heat transfer, however, involves sources \emph{within} one of the scatterers, preventing a simple optical theorem.

To circumvent this issue we reframe the scattering problem (without approximation). We define the ``incident'' field to be the unknown field emanating from body 1, and the ``scattered'' field to arise only with the introduction of body 2. For a Green's function (GF) $\tens{G}^1$ that is the field of dipole in the presence of body 1, the fields are given by a standard integral-equation separation~\cite{Chew1995}, $\vect{E}_{\textrm{inc},1} = (i/\varepsilon_0\omega) \int_{V_1} \tens{G}_1 \vect{J}$ and  $\vect{E}_{\textrm{scat},1} = \int_{V_2} \tens{G}_1 \vect{P}$, where $\vect{J}$ are the stochastic source currents in body 1, $\vect{P}$ is the polarization field induced in body 2, and $\varepsilon_0$ is the vacuum permittivity. This decomposition permits an optimal theorem with respect to body 2, such that its extinction is proportional to $\Im \int_{V_2} \cc{\vect{E}_{\textrm{inc},1}} \cdot \vect{P}$ (its absorption~\cite{Jackson1999} is proportional to $\int_{V_2} \left|\vect{P}\right|^2$). The energy-conservation arguments from above imply that absorption in body 2 is bounded,
\begin{align}
    P_{\rm abs,2} &\leq \frac{\varepsilon_0 \omega}{2} \frac{|\chi_2(\omega)|^2}{\Im \chi_2(\omega)} \int_{V_2} |\vect{E}_{\rm inc,1}(\vect{x}_2)|^2,
    \label{eq:abs_v2_limit}
\end{align}
which is formally derived by variational calculus~\cite{Miller2015subm}. To achieve this limit, the optimal polarization field must be proportional to the incident field, $\vect{P} \sim \vect{E}_{\textrm{inc},1}$, to maximize the extinction overlap integral. In the near field, where source fields rapidly decay, negative-permittivity metals that support surface-plasmon modes can achieve this condition, as we will demonstrate. 

The limit in \eqref{abs_v2_limit} reduces the optimal-flux problem to a question of how large the emitted field $\vect{E}_{\textrm{inc},1}$ can be in $V_2$. Inserting $\vect{E}_{\textrm{inc},1}$ into \eqref{abs_v2_limit} yields an integral of the stochastic currents, which is determined by the fluctuation-dissipation theorem~\cite{Joulain2005}, $\left\langle J_j(\vect{x},\omega),\cc{J_k(\vect{x}',\omega)} \right\rangle = 4 \varepsilon_0 \omega \Theta(\omega,T_1) \Im \left[ \chi\left(\omega\right) \right] \delta_{jk} \delta(\vect{x} - \vect{x}') / \pi$, such that the ensemble-averaged emitted field at $\vect{x}_2$ in $V_2$ is $\left\langle \left|\vect{E}_{\textrm{inc},1}(\vect{x}_2)\right|^2 \right\rangle =  4\varepsilon_0 \omega \Theta \left(\Im \chi_1\right) \int_{V_1} \left\|\tens{G}_{1}(\vect{x}_2,\vect{x}_1)\right\|_F^2$, where $\left\|\cdot \right\|_F$ denotes the Frobenius norm~\cite{Trefethen1997}. By reciprocity~\cite{Rothwell2001} one can exchange the positions in the integrand, $\vect{x}_1 \leftrightarrow \vect{x}_2$ (while transposing the GF, but the transpose does not affect the norm), such that emission from $V_1$ is equivalent to \emph{absorption} for free-space sources in $V_2$, as in \figref{fig1}(c). Absorption is bounded by energy conservation~\cite{Miller2015subm}, limiting the emitted-field magnitude:
\begin{align}
    \left\langle \left|\vect{E}_{\textrm{inc},1}(\vect{x}_2)\right|^2 \right\rangle \leq 4\varepsilon_0 \omega \Theta \frac{\left|\chi_1\right|^2}{\Im \chi_1} \int_{V_1} \left\|\tens{G}_{0}(\vect{x}_1,\vect{x}_2)\right\|_F^2
    \label{eq:einc_bound}
\end{align}
where $\tens{G}_0$ is the \emph{free-space} GF, cf. \figref{fig1}(d). Inserting \eqref{einc_bound} into \eqref{abs_v2_limit} and separating the Planck spectrum by \eqref{heat_transfer}, the maximum flux between two bodies is
\begin{align}
    \Phi(\omega) \leq \frac{2}{\pi} \frac{\left|\chi_1(\omega)\right|^2}{\Im \chi_1(\omega)} \frac{\left|\chi_2(\omega)\right|^2}{\Im \chi_2(\omega)} \int_{V_1} \int_{V_2} \left\| \tens{G}_0(\vect{x}_1, \vect{x}_2)\right\|_F^2.
    \label{eq:phi_lim_int}
\end{align}
The limit of \eqref{phi_lim_int} can be further simplified. In the near field, $\tens{G}_0$ is ideally dominated by the quasistatic term $\sim$~$1/r^3$, which is primarily responsible for the evanescent waves that enable greater-than-black-body heat-transfer rates~\cite{Joulain2005,Basu2009}. Dropping higher-order terms (further discussed in~\footnote{see Supplementary Material [url], which includes Refs.~\onlinecite{Cardona1971,Kong1975}}), we bound \eqref{phi_lim_int} by integrating over the infinite half-spaces containing $V_1$ and $V_2$, assuming a separating plane between the two bodies. (If not, e.g. between two curved surfaces, only the coefficients change.) For bodies separated by a distance $d$, the integral over the (infinite) area $A$ is given by~\cite{Note1} $\int_{V_1,V_2} \| \tens{G}_0 \|_F^2 = A / 32 \pi d^2$, yielding flux limits per area or relative to a black body with flux $\Phi_{\rm BB} = k^2 A / 4 \pi^2$~\cite{Joulain2005}:
\begin{align}
    \frac{\Phi(\omega)}{A} &\leq \frac{1}{16\pi^2 d^2} \frac{\left|\chi_1(\omega)\right|^2}{\Im \chi_1(\omega)} \frac{\left|\chi_2(\omega)\right|^2}{\Im \chi_2(\omega)}.
    \label{eq:phi_lim_area} \\
    \frac{\Phi(\omega)}{\Phi_{\rm BB}(\omega)} &\leq \frac{1}{4(kd)^2} \frac{\left|\chi_1(\omega)\right|^2}{\Im \chi_1(\omega)} \frac{\left|\chi_2(\omega)\right|^2}{\Im \chi_2(\omega)}.
    \label{eq:phi_lim_bb}
\end{align}

\eqrefrange{phi_lim_int}{phi_lim_bb} are fundamental limits to the near-field spectral heat flux between two bodies and form the central results of this Letter. They arise from basic limitations to the currents that can be excited in dissipative media, and their derivations further suggest physical characteristics of the optimal response in near-field heat transfer: an \emph{optimal emitter} enhances and absorbs near-field waves from reciprocal external sources \emph{in the absence of the absorber} whereas an optimal absorber enhances and absorbs near-field waves from the emitter, \emph{in the presence of the emitter}. These principles can be understood by working backwards through \figref{fig1}. The optimal-emitter condition identifies the largest field that can be generated in an exterior volume ($V_2$) by considering the reciprocal absorption problem, per \figref{fig1}(c). Reinserting the absorber, cf. \figref{fig1}(b), should not reflect the emitted field but rather enhance and absorb it. Because heat flux is symmetric with respect to absorber--emitter exchange, both bodies should satisfy each condition (induced currents proportional to source fields). \eqref{phi_lim_int} can be interpreted as sources throughout the emitter generating free-space dipolar fields $\tens{G}^0$ enhanced by $\left|\chi_1\right|^2 / \Im \chi_1$, which are further enhanced by $\left|\chi_2\right|^2 / \Im \chi_2$ and absorbed. The dipole--dipole interactions are bounded by their separation distance~\cite{Piestun2000,Miller2000a}, leading to simple shape-independent limits in \eqrefrange{phi_lim_int}{phi_lim_bb}. Ideal structures that achieve these limits can have significantly greater heat transfer than black bodies, even if their spectral flux has a narrow bandwidth. Whereas the heat transfer between black bodies in the far field is $H/A = \sigma_{\rm SB}T^4$, where $\sigma_{\rm SB}$ is the Stefan--Boltzmann constant~\cite{LienhardIV2011}, a straightforward calculation~cite{Note1} shows that ideal near-field heat exchange over a narrow bandwidth $\Delta \omega / \omega = \Im \chi / |\chi|$, typical of plasmonic systems~\cite{Wang2006a,Raman2013}, can achieve per-area transfer rates of
\begin{align}
    \frac{H}{A} \approx \sigma_{\rm SB} T^4 \frac{2}{7(kd)^2} \frac{|\chi|^3}{\Im \chi},
\end{align}
exhibiting both distance and material enhancements relative to the Stefan--Boltzmann rate.

The limits generalize~\cite{Note1} to local media with tensor susceptitbilities via the replacement $|\chi|^2 / \Im \chi \rightarrow \left\|\tens{\chi} \left(\Im \tens{\chi}\right)^{-1} \tens{\chi}^\dagger\right\|_2$. Nonlocal effects, which appear below 10nm length scales~\cite{Singer2015} and which regularize the $1/d^2$ divergence~\cite{Joulain2005}, are outside the scope of these limits, but we believe that a generalization to nonlocal $\tens{\chi}$ is possible and have preliminary results~\footnote{O. D. Miller et al., in progress.} suggesting that ``hydrodynamic''~\cite{Ciraci2012,Raza2015} nonlocal materials cannot not surpass the local-$\tens{\chi}$ bounds.

\begin{figure}[t!]
    \includegraphics[width=\linewidth]{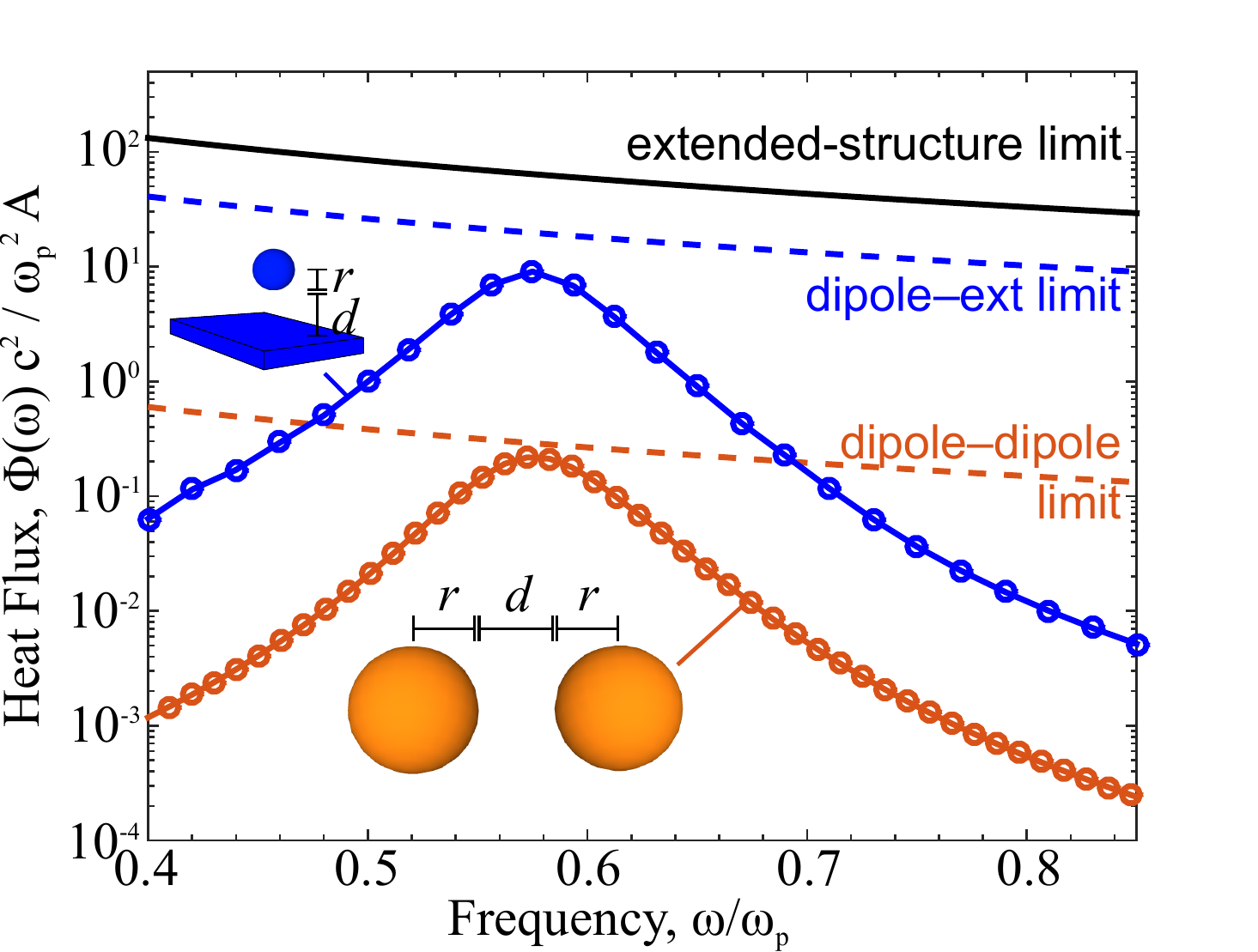}
    \caption{\label{fig:fig2} Comparison of heat flux in sphere--sphere and sphere--plate structures to the analytical limits of \eqreftwo{phi_lim_dip_dip}{phi_lim_dip_ext}. Two Drude metal spheres (orange circles, fit to a solid line) approach the dipole--dipole limit (dashed orange) at their resonant frequency, $\omega_{\rm res} \approx \omega_p / \sqrt{3}$. A sphere and a plate (blue circles) approach within a factor of two of the limit between dipolar and extended objects (dashed blue), if the material resonance of the plate is slightly modified (see text). In each case the separation is $d = 0.1c / \omega_{\rm res}$, with sphere radii $r = d/5$. The flux rates exhibit the material enhancement factor $|\chi|^4 / (\Im \chi)^2$, but not the near-field enhancement factor, due to the lack of large-area interactions. The sphere area $A$ is taken to be the cross-section $\pi r^2$.}
\end{figure}

\emph{Dipolar Interactions---}If one of the bodies is small enough for its response to be dipolar, the optimal-absorber and optimal-emitter conditions converge: the polarization currents induced in each structure by free-space dipoles in place of the opposite structure must be proportional to the incident fields. This condition is satisfied for two-dipole transfer, and the enhancement of the emitted and absorbed fields is possible via ``plasmonic'' resonances in metallic nanoparticles. For two identical particles with volumes $V$, tip-to-center-of-mass distances $r$, and tip-to-tip separation $d$, \eqref{phi_lim_int} limits the flux:
\begin{align}
    \left[\Phi(\omega)\right]_{\textrm{dipole--dipole}} \leq \frac{3}{4\pi^3} \frac{\left|\chi_1(\omega)\right|^2}{\Im \chi_1(\omega)} \frac{\left|\chi_2(\omega)\right|^2}{\Im \chi_2(\omega)} \frac{V^2}{\left(2r+d\right)^6}.
    \label{eq:phi_lim_dip_dip}
\end{align}
The radiative flux between quasistatic metal spheres is known analytically~\cite{Joulain2005} and peaks at the limit given by \eqref{phi_lim_dip_dip}.
\begin{figure*}
    \includegraphics[width=\textwidth]{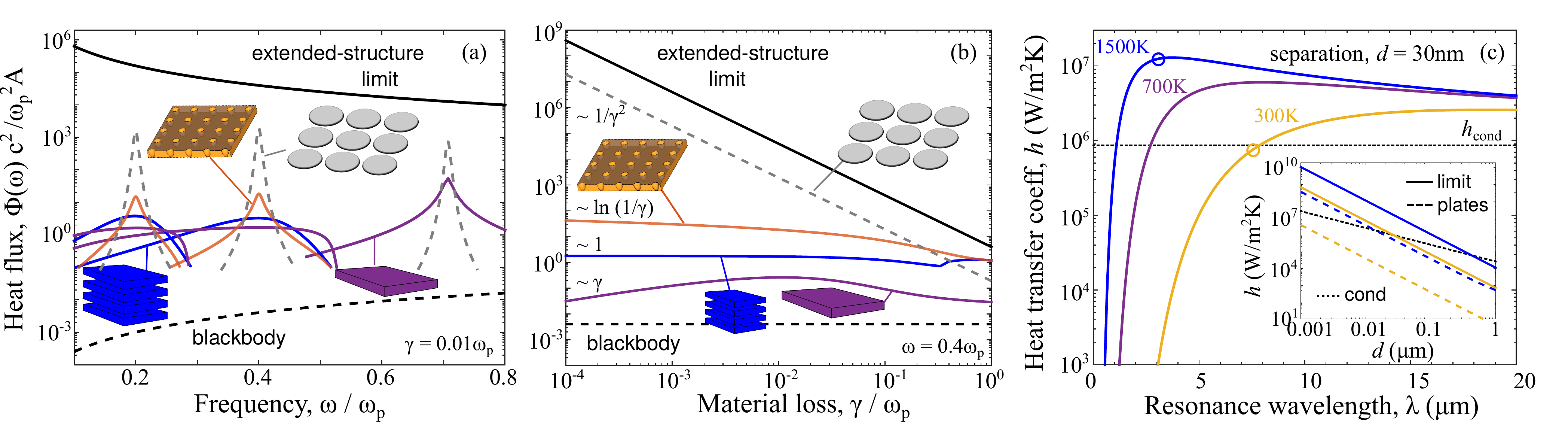}
    \caption{\label{fig:fig3}(a,b) Comparison of heat flux between mirror images of large-area Drude-metal structures separated by $d=0.1c/\omega_p$. (a) Structures optimized for maximum flux at three frequencies, $\omega = (0.2,0.4,1/\sqrt{2})\omega_p$, for a material loss rate $\gamma = 0.01\omega_p$. Thin films (purple), hyperbolic metamaterials (blue), and elliptical metamaterials (orange) exceed black-body enhancements but fall far short of the limit (black) from \eqref{phi_lim_area}. The dashed silver line represents the heat transfer for an idealized plasmonic-particle array without multiple scattering. (b) Optimized structures as a function of loss rate, for $\omega = 0.4\omega_p$. Each structure exhibits the $1/d^2$ near-field enhancement factor, but only the idealized particle array exhibits the $|\chi|^4 / (\Im \chi)^2 \sim 1/\gamma^2$ material enhancement factor. (c) Frequency-integrated heat transfer coefficient of a structure that reaches the single-frequency limit in \eqref{phi_lim_area} over a narrow bandwidth $\Delta\omega \propto \gamma$. Radiative heat exchange in this limit shows the possibility of surpassing conductive heat transfer through air (dotted) at $T=300$K (gold), which is not possible for plate--plate configurations (inset, dashed), and of significant further enhancements at higher temperatures (blue, purple).}
\end{figure*}

Heat transfer between a dipole and an extended structure is limited by integrating over the half-space occupied by any extended structure, yielding a maximum flux
\begin{align}
    \left[\Phi(\omega)\right]_{\textrm{dipole-to-ext}} \leq \frac{1}{8\pi^2} \frac{\left|\chi_1(\omega)\right|^2}{\Im \chi_1(\omega)} \frac{\left|\chi_2(\omega)\right|^2}{\Im \chi_2(\omega)} \frac{V}{(r+d)^3},
    \label{eq:phi_lim_dip_ext}
\end{align}
where $r+d$ is the distance between the extended structure and the particle's center. Heat flux between a sphere and a bulk metal, each supporting a plasmonic mode, can achieve half of the maximum flux~\cite{Mulet2001,Joulain2005,Note1} if the resonances align. This geometry falls short by a factor of two because planar surface plasmons exist only for TM polarization~\cite{Maier2007}, and thus the planar structure reflects near-field TE-polarized light emitted by the sphere. Neither structure exhibits the $1/d^2$ enhancement factor, which for dipolar coupling ($\sim$ $1/d^6$) requires interactions over two extended areas.

\figref{fig2} compares flux rates for sphere--sphere (orange circles) and sphere--plate (blue circles) geometries, computed by the fluctuating-surface current method~\cite{ReidScuffEM,Rodriguez2012,Rodriguez2013}, to the limits of \eqreftwo{phi_lim_dip_dip}{phi_lim_dip_ext} (orange and blue dashed lines, resp.). The spheres are modeled by Drude susceptibilities~\cite{Jackson1999} with plasma frequency $\omega_p$ and loss rate $\gamma = 0.1\omega_p.$ The ``plate'' is simulated by a very large ellipsoid (volume $\approx$ 7000$\times$ larger than the sphere) comprising a material with a modified plasma frequency, $\omega_{p,\text{pl}} = \sqrt{2/3} \omega_p$, and a modified loss rate, $\gamma_\text{pl} = 2\gamma / 3$, to align the resonant frequencies of the sphere and plate without modifying the flux limit. In each case the separation distance $d = 0.1 c / \omega_{\rm res}$ and the sphere radii are $r = d/5$. The computations support the analytical result that the dipolar limits can be approached to within at least a factor of two.

\emph{Extended Structures---}For extended structures that do not behave like single dipoles, the optimal-absorber constraint is more demanding in that the absorber should enhance the emitted field while accounting for interactions between the two bodies. We will show that common planar structures do not exhibit this behavior but that nanostructured media offer the possibility of approaching it.

Bulk metals (negative-permittivity materials) support surface plasmons that enable greater-than-blackbody heat flux at their resonant frequency. Individually, a single metal interface nearly satisfies the optimal-emitter condition, emitting near-field waves over a broad bandwidth of surface-parallel wavevectors (which enabled the nearly optimal sphere--plate transfer above). However, when a second metal is brought close to the first, it \emph{reflects} most of the incident field, except over a narrow wavevector-bandwidth, due to multiple-scattering effects between the bodies. The failure of the two-metal geometry to achieve the optimal-absorber condition leads to a peak spectral heat flux, at the surface-plasmon frequency $\omega_{\rm sp}$, of approximately~cite{Note1}
\begin{align}
    \left[ \frac{\Phi(\omega_{\rm sp})}{A} \right]_{\textrm{plate-to-plate}} = \frac{1}{4\pi^2 d^2} \ln \left[ \frac{|\chi|^4}{4(\Im \chi)^2} \right]
    \label{eq:phi_plate_plate}
\end{align}
which is significantly smaller than the limit in \eqref{phi_lim_area} due to the weak, logarithmic material enhancement. \eqref{phi_plate_plate} appears to be new and is a significantly better approximation than planar bounds that do not account for material loss~\cite{Pendry1999,Ben-Abdallah2010a}, as discussed in the SM~cite{Note1}. The shortcomings of the bulk-metal interactions cannot be overcome with simple metamaterial or thin-film geometries. The flux rate between hyperbolic metamaterials (HMMs) is material-independent~\cite{Biehs2012,Note1}. Optimal thin films behave similarly to HMMs~\cite{Miller2014a}, thereby also falling short of the limits. ``Elliptical'' metamaterials, with nearly isotropic effective permittivities, exhibit resonances for $\chi_{\rm eff} \approx -2$ and thus transfer heat at a rate similar to \eqref{phi_plate_plate}, limited by the same interference effects discussed above, and because $|\chi_{\rm eff}|^4 \ll |\chi|^4$.

\figref{fig3}(a,b) demonstrates the shortcomings of such structures, showing the computed heat flux between mirror images of thin-film (purple), hyperbolic-metamaterial (blue), and elliptical-metamaterial (orange) structures, as a function of (a) frequency and (b) material-loss rate, for a fixed separation $d=0.1c/\omega_p$. Assuming smooth surfaces without roughness, the structural parameters are computationally optimized~cite{Note1} using a derivative-free local optimization algorithm~\cite{Powell1994,JohnsonNLOpt}. \figref{fig3}(b) shows that the sub-optimal performance can be attributed primarily to the fact that the structures do not exhibit the material enhancement factor $|\chi|^4 / \left(\Im \chi\right)^2 \sim 1 / \gamma^2$, as predicted by \eqref{phi_plate_plate} and due to the significant reflections in such geometries.

The spectral heat flux of the limit in \eqref{phi_lim_int} can be interpreted as the exchange of enhanced free-space dipole fields, as discussed above. Guided by this intuition, we include in \figref{fig3}(a,b) the heat flux between close-packed arrays of oblate disk ellipsoids (dashed silver lines), small enough to be dipolar. We idealize their response as the additive sum of \eqref{phi_lim_dip_dip} over a lattice neglecting multiple scattering (i.e. in a Born approximation)~\cite{Phan2013} and accounting for the polarization-dependence of non-spherical ellipsoids~\cite{Bohren1983}. This structure combines the individual-particle interactions that exhibit the material enhancement (which planar bodies do not) with the large-area interactions that exhibit $1/d^2$ near-field enhancement (which isolated bodies do not). \figref{fig3}(a,b) suggest the possibility for two to three orders of magnitude enhancement by periodic structuring and tailored local interactions.

Experimental measurements of radiative heat transfer are done in vacuum~\cite{Hu2008,Rousseau2009,Ottens2011} because radiative transfer is dominated by conductive transfer through an air gap. Achieving the limits presented here, even over a narrow bandwidth, could transform this landscape. \figref{fig3}(c) shows the heat-transfer coefficient $h = \int \Phi (\partial \Theta / \partial T) {\rm d}\omega$ for extended Drude-metal structure with loss rates $\gamma = 0.01\omega_p$ (appropriate e.g. for Ag and Au~\cite{Palik1998}). For Lorentzian-shaped energy transfer with tunable center frequency $\omega_{\rm res} = \omega_p / \sqrt{2}$, peaked at the limit given by \eqref{phi_lim_area}, with a bandwidth $\Delta\omega = \gamma$~\cite{Wang2006a,Raman2013,Note1}, radiative transfer can surpass conductive (thermal conductivity $\kappa_{\rm air} = 0.026$W/m$\cdot$K~\cite{Haynes2013}) even at $T=300$K. In the inset we fix the wavelengths at $\lambda=7.6\mu$m for $T=300$K and $\lambda=3\mu$m for $T=1500K$, and plot $h$ as a function of distance for plate--plate (dashed) and optimal (solid) transfer. We find that radiative transfer can surpass conductive at separation of $d=50$nm at $300$K and almost $d=0.5\mu$m at $T=1500$K, gap sizes that are readily achievable in experiments. 

Radiative heat transfer at the nanoscale is a nascent but growing field. Calculations have primarily been for dipolar~\cite{Pendry1999,Mulet2001,Volokitin2007} or highly symmetric bodies~\cite{Loomis1994,Pendry1999,Xu1994,Fu2006,Hu2008,Ottens2011,VanZwol2011,Biehs2007,Francoeur2008,Francoeur2009,Ben-Abdallah2009,Francoeur2010,Basu2011,Miller2014a,Biehs2011,Francoeur2011,Joulain2010,Biehs2012,Biehs2013,Guo2013,Biehs2013,Otey2011,Perez-Madrid2013,Zheng2015}, with computational study of more complex geometries possible only recently~\cite{Rodriguez2012,Rodriguez2013,McCauley2012,Rodriguez2011,Perez-Madrid2008,Otey2014}. We have show that, guided by the physical principles presented here, a targeted search through the mostly uncharted near-field design space offers the prospect of orders-of-magnitude enhancements in radiative energy transfer.

\begin{acknowledgments}
We thank Athanasios Polimeridis for helpful discussions. ODM and SGJ were supported by the Army Research Office through the Institute for Soldier Nanotechnologies under Contract No. W911NF-07-D0004, and by the AFOSR Multidisciplinary Research Program of the University Research Initiative (MURI) for Complex and Robust On-chip Nanophotonics under Grant No. FA9550-09-1-0704. AWR was supported by the National Science Foundation under Grant No. DMR-1454836.
\end{acknowledgments}

\bibliography{/home/odmiller/texmf/bibtex/bib/library}
\end{document}


\title{Supplementary Materials: Shape-independent limits to near-field radiative heat transfer}

\author{Owen D. Miller}
\affiliation{Department of Mathematics, Massachusetts Institute of Technology, Cambridge, MA 02139}
\author{Steven G. Johnson}
\affiliation{Department of Mathematics, Massachusetts Institute of Technology, Cambridge, MA 02139}
\author{Alejandro W. Rodriguez}
\affiliation{Department of Electrical Engineering, Princeton University, Princeton, NJ 08544}

\begin{abstract}
    We provide: (1) a derivation and discussion of the higher-order terms in the heat flux limits, which tend to be very small for near-field heat transfer, (2) a derivation of the peak heat flux between two films, for a given material, (3) a derivation of the limits for a very general class of materials, (4) a derivation of the radiative heat transfer coefficient if the limiting flux rates are achieved, alongside a comparison to conductive heat transfer through air, and (5) an estimate of the frequency-integrated heat transfer for a narrow-band resonance.
\end{abstract}
\maketitle

\date{\today}

\section{Evaluation of integral limits and higher-order terms}

In this section, we present calculations and clarify the step needed
to go from Eq.~(5) to Eq.~(6) of the main text. Specifically, Eq.~(5)
is an integral bound that applies to any near- or far-field
interactions, depending only on conservation of energy
arguments. Eq.~(6) simplifies the bound for the case of near-field
heat transfer by assuming that the near-field quasistatic $1/r^3$ term
in $\tens{G}_0$ is the dominant term and integrating over the infinite
half-spaces occupied by the two bodies. (All equations and figures in this
Supplementary Material are preceded with an ``S,'' whereas
equations and figures without an ``S'' refer to the main
text.) Here we justify dropping the $1/r^2$ and $1/r$ terms in the
Green's function. Although for many structures it is known that
optimal near-field heat transfer is governed by high-wavevector waves
corresponding to the $1/r^3$ term, the mathematical justification for
dropping the terms is somewhat subtle. Integrated over infinite
half-spaces, the two terms diverge. We show that this divergence is
unphysical---originating from the optimal variational fields that are
appropriate in the near field but which do not satisfy Maxwell's equations
in the far field. Moreover, we show that for finite, reasonable
interaction distances, their contributions are negligible compared to
the contribution of the $1/r^3$ term. As shown in the text, the limit
of Eq.~(5), keeping only the $1/r^3$ term, yields very good agreement
with the response of sphere--sphere and sphere--plate interactions.

The squared Frobenius norm of the homogeneous Green's
function is:
\begin{align}
    \left\|\tens{G}_0\right\|_F^2 = \frac{k^6}{8\pi^2} \left[\frac{3}{\left(kr\right)^6} + \frac{1}{\left(kr\right)^4} + \frac{1}{\left(kr\right)^2} \right]
    \label{eq:GF0_sq}
\end{align}
which has contributions from $1/r^6$, $1/r^4$, and $1/r^2$ terms. For
convenience, instead of taking infinite half-spaces, we assume that
both bodies are contained within a circular cylinder of radius $R$ and
height $L$. The integral of the norm over both volumes is a
six-dimensional integral, but we bound it above by fixing the source
in one body at its center ($x=y=0$), and multiplying by the
cylindrical area $A = \pi R^2$:
\begin{align}
    \int_{V_1,V_2} \left\|\tens{G}_0\right\|_F^2 \leq A \int dz_1 \int dz_2 \int d\rho 2\pi \rho \left\|\tens{G}_0\right\|_F^2
\end{align}
where we have further simplified the integral using cylindrical coordinates. The multiplication by $A$ is exact for (infinitely wide) structures with translational and rotational symmetry; since we are interested in global bounds encompassing large structures it is thus a good approximation. The bound in Eq.~(6) of the main text comes from the $1/r^6$ term in the GF for an infinite volume (it is very weakly decreased for large but finite structures). The integral is given by:
\begin{align}
    \int_{V_1',V_2'} \frac{3}{r^6} = \frac{\pi A}{8 d^2},
\end{align}
where $V_1'$ and $V_2'$ are the infinite half-spaces containing the bodies. Multiplying by the prefactors in \eqref{GF0_sq} yields the bound in Eq.~(6) of the main text. Over finite volumes, the second term is more complicated:
\begin{align}
    \int_{V_1,V_2} \frac{1}{r^4} &= \pi A \left[ \log\left[ \frac{(d+L)^2}{d(d+2L)} \right] - \frac{2L+d}{R} \tan^{-1}\left(\frac{2L+d}{R}\right) \right. \nonumber \\
                                 &+ 2 \frac{L+d}{R} \tan^{-1}\left( \frac{L+d}{R} \right) \nonumber \\
                                 &+ \left. \frac{1}{2} \log\left[ \frac{\left((2L+d)^2 + R^2\right)\left(d^2 + R^2\right)}{\left((L+d)^2 + R^2\right)^2} \right] - \frac{d}{R} \tan^{-1}\left(\frac{d}{R}\right) \right]
    \label{eq:second_term}
\end{align}
The third term is given by:
\begin{align}
    \int_{V_1,V_2} \frac{1}{r^2} &= \pi A \left[ \frac{R^2}{2} \log\left[ \frac{\left((L+d)^2 + R^2\right)^2}{\left((2L+d)^2 + R^2\right) \left(R^2 + d^2\right)} \right] \right. \nonumber \\
                                 &+ \frac{(2L+d)^2}{2} \log\left[1 + \frac{R^2}{(2L + d)^2} \right] \nonumber \\
                                 &- \left(L + d\right)^2 \log\left[1 + \frac{R^2}{(L + d)^2}\right] + \frac{d^2}{2} \log\left(1 + R^2 / d^2\right) \nonumber \\
                                 &+ 2 R \left(d + 2L\right) \tan^{-1}\left(\frac{d+2L}{R}\right) \nonumber \\
                                 &- \left. 4R\left(d + L\right) \tan^{-1}\left(\frac{d+L}{R}\right) + 2Rd \tan^{-1}\left(\frac{d}{R}\right) \right]
    \label{eq:third_term}
\end{align}
\eqreftwo{second_term}{third_term} are difficult to disentangle so we consider large but finite volumes. Large bodies satisfy 
\begin{align}
    L,R \gg d
\end{align}
such that their sizes are much larger than their spacing. Not only do
$L$ and $R$ represent the physical sizes of the bodies, they also
represent the interaction sizes: they are the volumes over which
polarization currents within the respective bodies transfer
energy. Near-field interactions by definition occur between charges or
currents at the subwavelength scale, such that one is typically
interested in sizes $L \ll \lambda$. Conversely, surface waves between
structures are example of coherent subwavelength interactions that
potentially take place over distances much greater than the
wavelength, $R \gg \lambda$. Thus the finite-but-large asymptotic
expansion relevent for near-field heat transfer can be made by taking
\begin{align}
    d \ll L \ll R
\end{align}
for the two circular cylinders with radii $R$, heights $L$, and separation distance $d$. In this asymptotic limit, the terms simplify:
\begin{align}
    \frac{1}{\pi A} \int_{V_1',V_2'} \frac{3}{r^6} &= \frac{1}{8 d^2} \\
    \frac{1}{\pi A} \int_{V_1,V_2} \frac{1}{r^4} &\approx \log\left(\frac{L}{2d}\right) \\
    \frac{1}{\pi A} \int_{V_1,V_2} \frac{1}{r^2} &\approx 2L^2 \log\left(\frac{R}{4L}\right)
\end{align}
The divergences in the second and third terms are relatively weak. The
second term is negligible compared to the third term, which tends to
be very small compared to the first. The comparison between the first
and third term essentially compares $1/(kd)^2$ versus $(kL)^2$; even
in a generous upper bound in which $kL \approx 1$, the third term is
still much smaller than $1/(kd)^2 \gg 1$. In Table 1 we compare the
bound arising from Eq.~(6) to the bound that would arise from adding
\eqreftwo{second_term}{third_term} to Eq.~(6). We see that for
near-field distances ($d \ll \lambda$), even very large estimates of
the interaction distances $L$ and $R$ lead to only small modifications
to the upper limit, on the order of $1\%$ and in some cases
significantly smaller.

\begin{center}
\begin{tabular}[c]{|c|c|c|c|c|c|}
    \hline
    kd & kL & kR & Eq.~(6) & Eq.~(6)+\eqreftwo{second_term}{third_term} & Rel. Error \\ \hline
    0.01 & 1 & 1 & 1250 & 1252 & 0.17$\%$ \\ \hline
    0.01 & 1 & 10 & 1250 & 1254 & 0.35$\%$ \\ \hline
    0.01 & 1 & 100 & 1250 & 1256 & 0.53$\%$ \\ \hline
    0.001 & 1 & 100 & $1.25\times10^5$ & $1.25008\times 10^5$ & 0.0063$\%$ \\ \hline
    0.001 & 10 & 1000 & $1.25\times10^5$ & $1.255\times 10^5$ & 0.38$\%$ \\ \hline
\end{tabular}
\end{center}

Finally, we note that these divergences arise even for far-field
interactions, where they are clearly unphysical because finite
blackbody limits to the flux per unit area are well known. The
unphysical divergences arise from the assumption that the optimal
polarization fields are proportional to the incident
fields. Such a condition is ideal and achievable for the $1/r^3$
contribution of $\tens{G}_0$ that typically dominates near-field
transfer, but is unphysical for the more slowly decaying $1/r^2$ and
$1/r$ terms: a constant energy flux is maintained in a lossy medium
over large length scales, which is physically impossible. One approach
would be to ``split'' the problem into near- and far-field
contributions, and to bound the interactions separately. However,
given the relatively weak nature of these contributions for finite
interaction distances ($< 1\%$), they can be ignored for near-field
radiative heat transfer, justifying the use of Eq.~(6) in the main
text.


\section{Heat transfer between bulk planar media}

\begin{figure}
\includegraphics[width=0.9\linewidth]{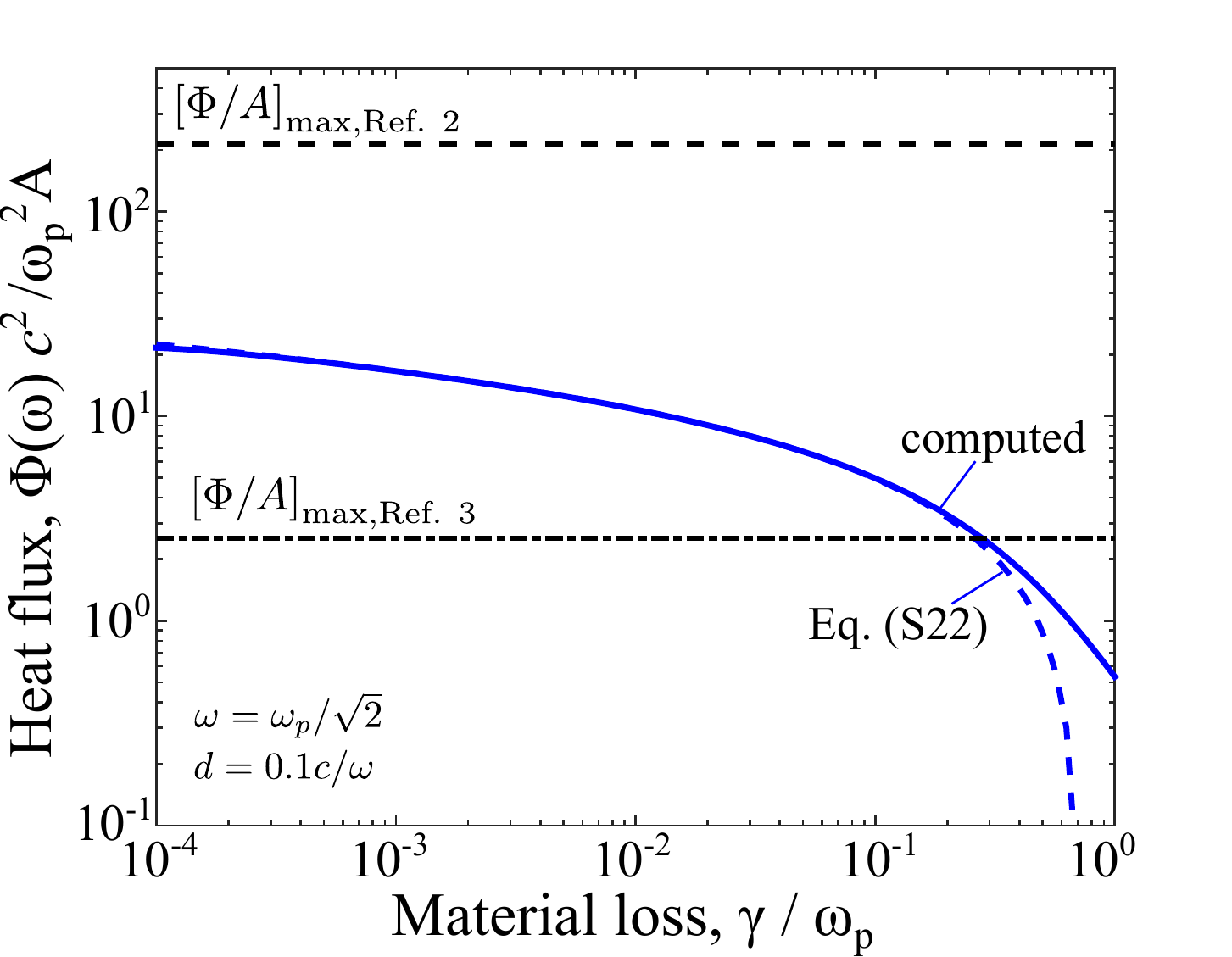}
\caption{\label{fig:figS1}Heat flux per unit area of two Drude-metal bulk media as a function of material loss rate, $\gamma / \omega_p$, at the resonant frequency $\omega=\omega_p/\sqrt{2}$ and at a fixed separation of $d=0.1c/\omega$. Except for very large loss, the heat flux approaches the approximate rate of \eqref{asy_exp}, confirming the logarithmic dependence on the material loss rate. Conversely, the limit of \citeasnoun{Pendry1999} is overly optimistic, and the ``limit'' of \citeasnoun{Ben-Abdallah2010a} is overly pessimistic. For the interatomic spacing that enters the limit of \citeasnoun{Pendry1999}, we took $b/\lambda \approx 1/1000$, which is appropriate e.g. for silver.}
\end{figure}

We derive the optimal heat-transfer rate between two planar bodies
comprising a material of susceptibility $\chi(\omega)$, corresponding
to Eq.~(10) of the main text. \citeasnoun{Basu2011} assumed a 
frequency-independent susceptibility, which they optimized for maximum 
heat transfer, whereas we assume a fixed (possibly frequency-dependent)
susceptibility. \citeasnoun{Pendry1999} and \citeasnoun{Ben-Abdallah2010a} 
also provide expressions for optimal heat flux between planar bodies, but 
their limits require wavevector-dependent material properties. The limits 
in both \citeasnoun{Pendry1999} and \citeasnoun{Ben-Abdallah2010a} arise
only because a finite maximum surface-parallel wavevector magnitude 
($k_\parallel$) is postulated: in \citeasnoun{Pendry1999} the maximum 
$k_{\parallel,\textrm{max}} = 1/b$ is chosen, where $b$ is the 
interatomic spacing of the metal; 
in \citeasnoun{Ben-Abdallah2010a}, the maximum $k_\parallel$ is inversely proportional 
to the gap spacing $d$, which does not account for large wavevectors 
that are possible when material losses are small. Although the interatomic 
spacing certainly sets an upper bound to the process as described by 
bulk materials, for lossy materials the loss is the limiting factor, not the 
interatomic spacing. We find a logarithmic dependence (and divergence) of 
the heat flux with the material loss rate, which we validate in \figref{figS1}.

The radiative heat flux $\Phi(\omega)$ between two planar slabs is
given by~\cite{Polder1971}
\begin{align}
    \Phi(\omega) = \frac{A}{4\pi^2} \int_0^\infty {\rm d}k_\parallel \, k_\parallel \left(T_p + T_s\right)
\end{align}
where $A$ is the area of the plates, $k_\parallel$ is the magnitude of the
surface-parallel part of the wavevector, and $T_s$ and
$T_p$ represent the field transmissions from slab 1 to slab 2
for $s$ and $p$ polarizations,
respectively. By symmetry, the surface-parallel
wavevector $k_\parallel$ is a conserved quantity between plane waves
in each medium. The heat flux is characterized by a strong
peak at a single $k_\parallel$ (for a given $\omega$) corresponding to
the metal-insulator-metal plasmonic mode. We show that at a given
frequency, the bandwidth in $k_\parallel$ is approximately constant,
while the peak energy transmission scales logarithmically with the
inverse of the material loss rate.

In the near field, we can focus only on the $p$-polarized transmission
coeffient for evanescent waves with $k_\parallel > k_0$. Assuming two
slabs of the same material, with reflectivity $r$ for waves incident
from air, the transmission coefficient is~\cite{Polder1971}:
\begin{align}
    T_p = \frac{4 \left[\Im(r)\right]^2 e^{-2\gamma d}}{\left|1 - r^2 e^{-2\gamma d}\right|^2}
    \label{eq:T_par}
\end{align}
where $\gamma = k_\parallel \sqrt{1 - k_0^2 / k_\parallel^2} \approx
k_\parallel$, assuming $k_\parallel \gg k_0$. Without the denominator, \eqref{T_par} would yield a $|\chi|^4 / (\Im \chi)^2$ enhancement from the plasmon waves at each surface, manifested in the
poles of $\Im r_p$~\cite{Cardona1971}. However, at the small distances necessary to transfer energy, the denominator---heuristically originating from the infinite sum of reflected waves---has an identical pole that \emph{cancels} the one in the numerator. The resonances of $T_p$ are instead metal-insulator-metal modes, with energy levels split around the single-surface plasmon
energies~\cite{Maier2007}, as discussed in the main text. 

\citeasnoun{Pendry1999} and \citeasnoun{Ben-Abdallah2010a} find limits to the transfer by noting 
that at every $k_\parallel$ the maximum value of $T$ is 1 (note that for conventional 
metals such a tranmission would require a wavevector-dependent permittivity). 
They define $k_{p,\textrm{max}}^2=1/b^2$~\cite{Pendry1999} and 
$k_{p,\textrm{max}}^2=4/d^2$~\cite{Ben-Abdallah2010a}, respectively, 
yielding limits:
\begin{align}
    \left[\frac{\Phi(\omega)}{A}\right]_{\rm max,Ref.~2} &= \frac{1}{8\pi^2 b^2} \\
    \left[\frac{\Phi(\omega)}{A}\right]_{\rm max,Ref.~3} &= \frac{1}{2\pi^2 d^2}
\end{align}
for interatomic spacing $b$ and separation distance $d$.

Instead we seek a limit assuming a conventional (wavevector-independent) 
material susceptibility $\chi(\omega)$. Defining $x=2k_\parallel d$, the flux is given by:
\begin{align}
    \Phi(\omega) &= \frac{A}{4\pi^2 d^2} \int_0^\infty \frac{\left[\Im(r)\right]^2 x e^{-x}}{1 - 2\Re(r^2)e^{-x} + |r|^4 e^{-2x}} \,{\rm d}x \nonumber \\
                                       &= \frac{A}{4\pi^2 d^2} \int_0^\infty x f(x) \,{\rm d}x
\end{align}
where the integral lower bound can be set to zero because we have assumed $k_0 d \ll 1$, and $f(x)$ is defined by
\begin{align}
    f(x) = \frac{\left[\Im(r)\right]^2 e^{-x}}{1 - 2\Re(r^2)e^{-x} + |r|^4 e^{-2x}}.
    \label{eq:fdef}
\end{align}
At large $k_\parallel$, the reflectivity $r$ is approximately constant
and given by $r = (\varepsilon - 1) / (\varepsilon + 1)$. We will not
insert its exact form at the moment, but we will note that for the
optimal susceptibility (see below) the real part of $r$ is 0 and the
imaginary part is potentially large. It we define the average
[weighted by $f(x)$] value of $x$ as $x_0$, it follows that $\int x
f(x) = x_0 \int f(x)$ and hence $\Phi$ can be approximately given by:
\begin{align}
    \Phi(\omega) \approx \frac{x_0 A}{4\pi^2 d^2} \int_0^\infty f(x) \,{\rm d}x
\end{align}
The integral of $f$ can be worked out:
\begin{align}
    \int_0^\infty f(x) \,{\rm d}x &= \frac{\left[\Im(r)\right]^2}{\Im(r^2)} \left[\frac{\pi}{2} - \tan^{-1}\left( \frac{1 - \Re(r^2)}{\Im(r^2)} \right)\right] \nonumber \\
                                  &= \frac{\left[\Im(r)\right]^2}{\Im(r^2)} \tan^{-1} \left( \frac{\Im(r^2)}{1 - \Re(r^2)} \right) \nonumber \\
                                  &\approx \frac{\left[\Im(r)\right]^2}{1 - \Re(r^2)}
\end{align}
where we used $\tan^{-1}(1/x) = \pi/2 - \tan^{-1}(x)$, and for $x$ small, $\tan^{-1}(x) \approx x$. For the final step, we can write $\Re(r^2) = \left[\Re(r)\right]^2 - \left[\Im(r)\right]^2 = 1 - \left[\Im (r)\right]^2$. To find the value of $x_0$, we approximate it (verifying later) as the value of $x$ at which $f(x)$ peaks. Setting the derivative of $f$ in \eqref{fdef} to zero yields:
\begin{align}
    x_0 = \ln |r|^2.
\end{align}
Because $r = 1 / [1 + 2 / \chi(\omega)]$, the optimal frequency for
maximum $|r|$ is given by the frequency such that $\Re
(-1/\chi(\omega)) = 1/2$. At this frequency, $r = i |\chi|^2 / 2 \Im
\chi$ and we have:
\begin{align}
    &x_0 = \ln \left[ \frac{|\chi|^4}{4\left(\Im \chi\right)^2}\right] \\
    \int_0^\infty &f(x) = 1
\end{align}
Thus at the optimal frequency, maximum energy transmission occurs for 
$k_\parallel$ logarthmically proportional to the inverse of the
material loss rate, and the bandwidth in $k_\parallel$ is
constant. Hence, the radiative flux rate between the two slabs is
given by:
\begin{align}
    \frac{\Phi(\omega)}{A} \approx \frac{1}{4\pi^2 d^2} \ln \left[ \frac{|\chi|^4}{4\left(\Im \chi\right)^2}\right]
    \label{eq:asy_exp}
\end{align}
The asymptotic expression in \eqref{asy_exp} is almost identical to
the limit in Eq.~(10) in the main text, except that the flux rate
scales logarithmically instead of linearly with $|\chi|^4/\left(\Im
\chi\right)^2$.

Conversely, for hyperbolic metamaterials, the optimal near-field heat flux is~\cite{Biehs2012}
\begin{align}
    \left[ \frac{\Phi(\omega_{\rm res})}{A} \right]_{\textrm{HMM-to-HMM}} = \frac{\ln 2}{4\pi^2 d^2}.
    \label{eq:phi_hmm_hmm}
\end{align}
HMMs therefore do not exhibit any material enhancement; because the
resonant modes are inside the bulk rather than at the surface, there
is no divergence in the lossless limit.

\section{Limits for general media}
For clarity, and with regard to practical relevance,  we presented in the main text only limits to heat flux between nonmagnetic, isotropic bodies. Here we derive the limits for more general media, leading to the generalization $|\chi|^2 / \Im \chi \rightarrow \left\|\tens{\chi} \left(\Im \tens{\chi}\right)^{-1} \tens{\chi}^\dagger\right\|_2$, as discussed in the main text. For notational simplicity we define $\tens{\xi} = -\tens{\chi}^{-1}$, in which case the generalization is $|\chi|^2 / \Im \chi \rightarrow \left\| \left(\Im \tens{\xi}\right)^{-1} \right\|_2$, where $\|\cdot\|_2$ is the induced matrix 2-norm~\cite{Trefethen1997}. This generalization applies even for non-reciprocal media, thanks to a generalized reciprocity theorem~\cite{Kong1975}.

The Maxwell curl equations are
\begin{align}
    \nabla \times \vect{H} + i \omega \vect{D} &= \vect{J}_e \\
    -\nabla \times \vect{E} + i \omega \vect{B} &= \vect{J}_m
\end{align}
To simplify notation going forward, we will encapsulate electric and magnetic components of fields and currents into six-component vectors. We denote the fields by $\psi$, the free currents by $\sigma$, and the induced polarization currents by $\nu$:
\begin{align}
    \psi = \begin{pmatrix}
        \vect{E} \\
        \vect{H}
    \end{pmatrix} \quad
    \sigma = \begin{pmatrix}
        \vect{J}_e \\
        \vect{J}_m
    \end{pmatrix} \quad
    \nu = \begin{pmatrix}
        \vect{P} \\
        \vect{M}
    \end{pmatrix}
\end{align}
The polarization currents within a body are related to the internal fields by the 6$\times$6 tensor susceptibility $\tens{\chi}$,
\begin{align}
    \nu = \tens{\chi} \psi.
\end{align}
Given these definitions, the Maxwell curl equations can be rewritten:
\begin{align}
    \left[ 
        \begin{pmatrix}
            i\omega\varepsilon_0 & \nabla \times \\
            -\nabla \times & i \omega \mu_0
        \end{pmatrix}
    + i \omega \tens{\chi} \right] \psi = \sigma
\end{align}
Following the derivation in the main text, the first step is to define a Green's function (GF), $\tens{\Gamma}_1$, in the presence of only body 1:
\begin{align}
    \left[\begin{pmatrix}
        i\omega\varepsilon_0 & \nabla \times \\
        -\nabla \times & i \omega \mu_0
    \end{pmatrix} + i\omega \tens{\chi}_1 \right] \tens{\Gamma}_1(\vect{x},\vect{x}_0) = -i\omega \tens{I} \delta \left(\vect{x}-\vect{x}_0\right)
\end{align}
where it is implicit that $\tens{\chi}_1 = 0$ at points outside of $V_1$. Then the total fields in the presence of both bodies, excited by stochastic currents in body 1, satisfy the integral equation
\begin{align}
    \psi(\vect{x}) &= \frac{i}{\omega} \int_{V_1} \tens{\Gamma}_1(\vect{x},\vect{x}_0) \sigma(\vect{x}_0) + \int_{V_2} \tens{\Gamma}_1(\vect{x},\vect{x}_0) \tens{\chi}_2 \psi(\vect{x}_0) \\
                   &= \psi_{\textrm{inc},1} + \int_{V_2} \tens{\Gamma}_1(\vect{x},\vect{x}_0) \tens{\chi}_2 \psi(\vect{x}_0).
\end{align}
Now the fields incident from body 1 have been separated from the ``scattered'' fields that arise only from the introduction of body 2, while fully accounting for interactions between the two bodies. Then the powers absorbed and extinguished by body 2 are given by:
\begin{align}
    P_{\rm abs} &= \frac{\omega}{2} \Im \int_{V_2} \cc{\nu} \cdot \tens{\xi}_2 \nu \\
    P_{\rm ext} &= \frac{\omega}{2} \Im \int_{V_2} \cc{\psi_{\textrm{inc},1}} \cdot \nu
\end{align}
where
\begin{align}
    \tens{\xi}_2 = -\tens{\chi}_2^{-1}
\end{align}
Constraining $P_{\rm abs}<P_{\rm ext}$ yields a limit to the absorbed power:
\begin{align}
    P_{\rm abs} &\leq \frac{\omega}{2} \int_{V_2} \cc{\psi_{\textrm{inc},1}} \cdot \left(\Im \tens{\xi}_2\right)^{-1} \psi_{\textrm{inc},1} \\
                &\leq \frac{\omega}{2} \left\| \left(\Im \tens{\xi}_2\right)^{-1} \right\|_2 \int_{V_2} \left| \psi_{\textrm{inc},1} \right|^2 
\end{align}
where the second inequality follows from the definition of the induced matrix 2-norm, $\left\|\cdot\right\|$. We can write out the squared magnitude of the incident field:
\begin{align}
    \left| \psi_{\textrm{inc},1} \right|^2 = \frac{1}{\omega^2} \int_{V_1} \int_{V_1} \sigma^\dagger (\vect{x}_1) \tens{\Gamma}_1^\dagger(\vect{x},\vect{x}_1) \tens{\Gamma}_1(\vect{x},\vect{x_1'}) \sigma(\vect{x_1'})
    \label{eq:psi_sq}
\end{align}
The fluctuation-dissipation theorem dictates that the ensemble average of the current--current correlation function is
\begin{align}
    \left\langle \sigma(\vect{x_1'}) \sigma^\dagger(\vect{x}_1) \right\rangle = \frac{4}{\pi} \omega \left[\Im \tens{\chi}_1\right] \delta(\vect{x}_1 - \vect{x_1'}) \Theta(\omega,T_1)
    \label{eq:fdt}
\end{align}
Inserting \eqref{fdt} into \eqref{psi_sq} yields the limit to the energy flux into body 2 (the Planck factor separately multiplies the flux to give the total power):
\begin{align}
    \Phi(\omega) \leq \frac{2}{\pi} \left\| \left(\Im \tens{\xi}_2\right)^{-1} \right\|_2 \Tr \int_{V_1} \int_{V_2} \tens{\Gamma}_1(\vect{x}_1,\vect{x}_2) \left(\Im \tens{\chi}_1\right) \tens{\Gamma}_1^\dagger (\vect{x}_1, \vect{x}_2)
    \label{eq:phi_1}
\end{align}
The integrand in \eqref{phi_1} relates the fields in $V_2$, in empty space, from sources in $V_1$, within body 1. To find limits to this quantity, it would be useful to transpose the source and measurement positions in the Green's functions. Even if body 1 consists of a nonreciprocal material, it is possible to switch the source and receiver positions if the material susceptibility,
\begin{align}
    \tens{\chi} = 
    \begin{pmatrix}
        \tens{\chi}_{11} & \tens{\chi}_{12} \\
        \tens{\chi}_{21} & \tens{\chi}_{22}
    \end{pmatrix}
\end{align}
is simultaneously transformed to a \emph{complementary} medium~\cite{Kong1975},
\begin{align}
    \tens{\chi}_C &=
    \begin{pmatrix}
        \tens{\chi}^T_{11} & -\tens{\chi}^T_{21} \\
        -\tens{\chi}^T_{12} & \tens{\chi}^T_{22}
    \end{pmatrix} \\
    &= S \tens{\chi}^T S
    \label{eq:comp_mat}
\end{align}
where
\begin{align}
    S = \begin{pmatrix}
        \mathbb{I} & \\
                   & -\mathbb{I}
    \end{pmatrix}
\end{align}
and $\mathbb{I}$ is the 3$\times$3 identity matrix. Defining $\tens{\Gamma}_C$ as the Green's function in the presence of the complementary-medium body 1, the modified reciprocity relation~\cite{Kong1975} dictates:
\begin{align}
    \tens{\Gamma_1}(\vect{x}_1, \vect{x}_2) = S \tens{\Gamma}^T_C(\vect{x}_2, \vect{x}_1) S
    \label{eq:recip}
\end{align}
We can then perform a number of simplifications on the integrand in \eqref{phi_1}, including the trace operator and pulling the imaginary operator out front:
\begin{align*}
    \Im &\Tr \tens{\Gamma}_1(\vect{x}_1, \vect{x}_2) \tens{\chi}_1 \tens{\Gamma}^\dagger_1(\vect{x}_1, \vect{x}_2) \\
        &= \Im \Tr S \tens{\Gamma}^T_C(\vect{x}_2, \vect{x}_1) S \tens{\chi}_1 S \cc{\tens{\Gamma}}_C(\vect{x}_2, \vect{x}_1) S \\
              &= \Im \Tr S \tens{\Gamma}^T_C(\vect{x}_2, \vect{x}_1) \tens{\chi}_{1C}^T \cc{\tens{\Gamma}}_C(\vect{x}_2, \vect{x}_1) S \\
              &= \Im \Tr \tens{\Gamma}^T_C(\vect{x}_2, \vect{x}_1) \tens{\chi}_{1C}^T \cc{\tens{\Gamma}}_C(\vect{x}_2, \vect{x}_1) \\
              &= \Im \Tr \tens{\Gamma}^\dagger_C(\vect{x}_2, \vect{x}_1) \tens{\chi}_{1C} \tens{\Gamma}_C(\vect{x}_2, \vect{x}_1)
\end{align*}
where the first equality uses reciprocity as defined by \eqref{recip}, the second equality uses the definition of the complementary medium, \eqref{comp_mat}, the third equality uses $\Tr SXS = \Tr X$, by the definition of $S$, and the final equality takes the transpose of the matrix product inside the trace. After applying these transformations, \eqref{phi_1} now represents a new absorption problem: the absorption inside the \emph{complementary} version of body one due to sources in empty space in $V_2$. This absorption problem can be bounded just as the previous one was, by energy conservation, such that
\begin{align}
    \Im &\int_{V_1} \tens{\Gamma}^\dagger_C(\vect{x}_2, \vect{x}_1) \tens{\chi}_{1C} \tens{\Gamma}_C(\vect{x}_2, \vect{x}_1) \\
    &\leq \left\|\left(\Im \tens{\xi}_{1C}\right)^{-1} \right\|_2 \int_{V_1} \tens{\Gamma}^\dagger_0(\vect{x}_2, \vect{x}_1) \tens{\Gamma}_0(\vect{x}_2, \vect{x}_1) 
\end{align}
where $\tens{\Gamma}_0$ is the free-space Green's function and $\tens{\xi}_{1C} = -\tens{\chi}_{1C}^{-1}$. It turns out that the norm of the loss rate for the complementary material is equal to the norm of the loss rate of the original material:
\begin{align*}
    \left\| \left( \Im \tens{\xi}_{1C} \right)^{-1} \right\|_2 &= \left\| -\left(\Im \tens{\chi}_{1C}^{-1} \right)^{-1} \right\|_2 \\
                                                               &= \left\| -\left(\Im \left[S \tens{\chi}_1^T S\right]^{-1} \right)^{-1} \right\|_2 \\
                                                               &= \left\| -\left(\Im \left[ \tens{\chi}_1^T \right]^{-1} \right)^{-1} \right\|_2 \\
                                                               &= \left\| -\left(\Im \left[ \tens{\chi}_1 \right]^{-1} \right)^{-1} \right\|_2 \\
    &= \left\| \left(\Im \tens{\xi}_1 \right)^{-1} \right\|_2
\end{align*}
through repeated application of the facts that $S^{-1} = S^\dagger = S$ and that transposing a matrix does not affect its norm. Finally, we relate the trace of the integrand to the Frobenius norm of the Green's function:
\begin{align}
    \Tr \tens{\Gamma}^\dagger_0 \tens{\Gamma}_0 = \left\| \tens{\Gamma}_0 \right\|_F^2
\end{align}
to ultimately yield a flux limit:
\begin{align}
\Phi(\omega) \leq \frac{2}{\pi} \left\| \left(\Im \tens{\xi}_{1}\right)^{-1} \right\|_2 \left\| \left(\Im \tens{\xi}_2\right)^{-1} \right\|_2 \int_{V_1} \int_{V_2} \left\| \tens{\Gamma}_0 (\vect{x}_1, \vect{x}_2) \right\|_F^2
\end{align}
that is precisely the generalization of Eq.~(5) in the main text, for a wide class of materials. The limit could even be extended to inhomogeneous media, although the exact geometry would need to be specified to know the material loss rate everywhere. 

\section{Radiative vs conductive heat-transfer coefficients}
We compare radiative heat transfer to conductive heat transfer and
derive the equations used for the plots shown in Fig.~3(c). The total
radiative heat transfer between two bodies is given by Eq.~(1) in the
main text, $H = \int \Phi(\omega) \left[\Theta(\omega,T_1) -
  \Theta(\omega,T_2)\right]\,{\rm d}\omega$. For a small temperature
differential between the bodies, the conductance (heat transfer per
unit temperature) per area $A$ is termed the \emph{radiative heat transfer
  coefficient} and is given by
\begin{align}
    h_\text{rad} &= \frac{1}{A} \int \Phi(\omega) \frac{\partial \Theta}{\partial
      T} \,{\rm d}\omega = \frac{1}{A} k_B \int \Phi(\omega) f(\omega) \,{\rm
      d}\omega,
\end{align}
where
\begin{align}
    f(\omega) = \left(\frac{\hbar \omega}{k_B T}\right)^2 \frac{e^{\hbar\omega/k_BT}}{\left(e^{\hbar\omega/k_BT} - 1\right)^2}
\end{align}
When considering the limits to radiative heat transfer between
metallic objects, one can expect that the resonances will have
relatively small decay rates and thus that $\Phi$ will be very narrow,
and much sharper than the Boltmann-like distribution $f(\omega)$ in
the integrand. Thus we approximate $h$ by
\begin{align}
  h_\text{rad} \approx \frac{1}{A} k_B f(\omega_0) \int \Phi(\omega) d\omega.
\end{align}
We take the metal to be a Drude metal with susceptibility
$\chi(\omega) = -\omega_p^2 / (\omega^2 + i\gamma\omega)$, for
simplicity. Moreover, we assume that the absorption and emission of
each body is described by a single sharp Lorentzian, with a narrow
bandwidth (full-width at half-max) given by $\Delta \omega =
\gamma$~\cite{Wang2006a,Raman2013}. This is much narrower than e.g. the
plane--plane and metamaterial structures in Fig.~3(a,b) and is in line
with the resonant heat transfer between two spheres or between a
sphere and a plate, depicted in Fig.~2 of the main text. 
The integral over $\Phi$ is then
\begin{align}
    \int \Phi(\omega) \,{\rm d}\omega = \frac{\pi \gamma}{2} \Phi(\omega_0)
    \label{eq:phi_lor}
\end{align}
and thus the radiative heat transfer coefficient is given by:
\begin{align}
    h_\text{rad} \approx \frac{1}{2} \pi \gamma k_B f(\omega_0) \frac{\Phi(\omega_0)}{A}
\end{align}
The single-frequency limit to the flux per unit area is given by Eq.~(6) in the main text, repeated here for a Drude metal:
\begin{align}
    \frac{\Phi(\omega_0)}{A} \leq \frac{1}{16\pi^2 d^2} \frac{\omega_p^4}{\gamma^2 \omega_0^2}
\end{align}
Thus the limit to the radiative heat transfer coefficient is
\begin{align}
  h_\text{rad} \leq \frac{k_B \omega_0}{32 \pi d^2} \frac{\omega_p^4}{\gamma \omega_0^3} f(\omega_0)
\end{align}
From a design perspective, not each of the parameters in Eq.~(10) is a
free parameter. The choice of temperature, for example, sets the
optimal frequency (a blackbody at $300K$ has maximum emission at
$7.6\mu$m wavelength). Similarly, the factor $\omega_p / \omega$ is
limited by the optimal aspect ratio, and the factor $\gamma /
\omega_p$ is set by the material loss rate. Hence, it is convenient to
rewrite Eq.~(10) as
\begin{align}
    h_{\rm rad} \leq \frac{k_B^2 T}{\hbar} \left[ \frac{1}{32 \pi d^2} \frac{\omega_p^4}{\gamma \omega^3} g(\omega) \right]
\end{align}
where $g= x^3 e^x / (e^x - 1)^2$ for $x=\hbar \omega / k_B T$.

The thermal conductivity of air is~\cite{Haynes2013}:
\begin{align}
    \kappa_{\rm air} = 0.026 \frac{\mathrm{W}}{\mathrm{m} \cdot
      \mathrm{K}}
\end{align}
Across a gap of size $d$, the conductive heat transfer coefficient is given by
\begin{align}
    h_{\rm cond} = \frac{\kappa}{d}
\end{align}
$h_{\rm rad}$ and $h_{\rm cond}$ are plotted in Fig.~3(c) in the main
text for a variety of wavelengths and temperatures; also included are
radiative heat transfer coefficients for plane--plane configurations,
which fall short of the limits presented and require extremely small
separation distances to even reach the conductive heat transfer
coefficient.  

\section{Resonant heat transfer}
One can similarly calculate the approximate frequency-integrated heat transfer for a narrow-band spectral flux. The heat transfer is given by
\begin{align}
    H = \int_0^\infty \Phi(\omega) \Theta(\omega,T)
\end{align}
where we've taken one of the bodies at temperature $T$ to be much hotter than the other body (such that $\Theta_1 - \Theta_2 \approx \Theta_1$). For a sharp, resonant spectral flux centered at $\omega=\omega_0$, we can take $\Theta$ roughly fixed at its value at $\omega_0$, similar to the approximation of $f$ in Sec.~4. For a flux with Lorentzian lineshape of bandwidth $\Delta \omega$, the transfer per area is
\begin{align}
    \frac{H}{A} \approx \Theta(\omega_0,T) \int_0^\infty \frac{\Phi(\omega)}{A} \,{\rm d}\omega = \frac{\pi\Delta\omega}{2} \frac{\Phi(\omega_0)}{A} \Theta(\omega_0,T)
\end{align}
by \eqref{phi_lor}. For a spectral flux that peaks at the limit given in our manuscript, we have
\begin{align}
    \frac{\Phi(\omega_0)}{A} = \frac{1}{16\pi^2d^2} \frac{\left|\chi\right|^4}{\left(\Im \chi\right)^2}
\end{align}
where for simplicity we've taken $\chi_1=\chi_2=\chi$. The Planck distribution factor is given by $\Theta = \hbar\omega_0 / (e^x - 1)$, where $x=\hbar\omega_0/k_B T$. For typical plasmonic systems~\cite{Raman2013} the loss rate is proportional to the material loss $\Im \chi / |\chi|$, such that we can approximate $\Delta \omega \approx \omega_0 \left(\Im \chi\right) / \left|\chi\right|$. Then the heat transfer per unit area is
\begin{align}
    \frac{H}{A} &= \frac{k_B T}{32\pi c^2 (kd)^2} \frac{\left|\chi\right|^3}{\Im \chi} \omega^3 \frac{x}{e^x-1} \\
                &= \frac{\pi^2 \left(k_B T\right)^4}{4 h^3 c^2 (kd)^2} \frac{\left|\chi\right|^3}{\Im \chi} \frac{x^4}{e^x-1}
\end{align}
Two far-away black bodies exchange heat at a rate $H/A = \sigma_{\rm SB} T^4$, where $\sigma_{\rm SB}$ is the Stefan--Boltzmann constant:
\begin{align}
    \sigma_{\rm SB} = \frac{2\pi^5 k_B^4}{15c^2 h^3}.
\end{align}
We can rewrite the near-field transfer in terms of $\sigma_{\rm SB}$,
\begin{align}
    \frac{H}{A} = \sigma_{\rm SB} T^4 \left(\frac{15}{8\pi^3} \frac{x^4}{e^x-1}\right) \frac{1}{\left(kd\right)^2} \frac{\left|\chi\right|^3}{\Im \chi}.
\end{align}
The term in parenthesis is maximum for $\hbar\omega / k_B T = x \approx 3.9$, in which case the term itself is $0.28904\ldots \approx 2/7$, such that we can write
\begin{align}
    \frac{H}{A} \approx \sigma_{\rm SB} T^4 \frac{2}{7 \left(kd\right)^2} \frac{\left|\chi\right|^3}{\Im \chi},
\end{align}
which is precisely the Stefan--Boltzmann ray-optics limit, scaled up by the distance enhancement $1/(kd)^2$ and by the material enhancement $|\chi|^3 / \Im \chi$.

\bibliography{/home/odmiller/texmf/bibtex/bib/library}